\documentclass[11pt,fleqn]{article}
\usepackage{epsfig}
\usepackage{amsfonts}
\usepackage{times}
\usepackage{mathptm}
\newcommand{\Section}[1]%
{\section{#1}\setcounter{equation}{0}%
\setcounter{theorem}{0}}

{\par\noindent{\em #1:\ }}%
{~\rule{2mm}{2mm}\par\bigskip}

\newcommand{\ret}{\nonumber \\}
\newcommand{\beq}{\begin{equation}}
\newcommand{\eeq}{\end{equation}}
\newcommand{\bem}{\begin{displaymath}}
\newcommand{\eem}{\end{displaymath}}
\newcommand{\beqar}{\begin{eqnarray}}
\newcommand{\eeqar}{\end{eqnarray}}
\newcommand{\average}[1]{\langle #1 \rangle}
\newcommand{\abs}[1]{\left| #1 \right|}

\font\titlefnt=cmbx10 scaled \magstep2

\begin{document}
\mathindent 0mm
\newpage\thispagestyle{empty} 
\begin{flushright} HD--TVP--97--06\end{flushright}
\vspace*{2cm}
\begin{center} 
{\titlefnt Noise induced transport at zero temperature\\ \vspace*{0.4cm}
}
\vskip3cm
Heiner Kohler\footnote[1]
{E--mail: kohler@tphys.uni-heidelberg.de}\\ 
Andreas Mielke\footnote[2]
{E--mail: mielke@tphys.uni-heidelberg.de}\\ 
\vspace*{0.2cm}
Institut f\"ur Theoretische Physik,\\
Ruprecht--Karls--Universit\"at,\\
Philosophenweg 19, \\
D-69120~Heidelberg, F.R.~Germany
\\
\vspace*{1.5cm}
\today
\\
\vspace*{2cm}
\noindent

{\bf Abstract}
\end{center}

\vspace*{0.2cm}\noindent
We consider a particle in the over-damped regime at zero
temperature under the influence of a sawtooth potential and of a noisy force,
which is correlated in time. A current occurs, even if the mean of
the noisy force vanishes. We calculate the stationary probability distribution
and the stationary current. We discuss, how  these items depend on the
characteristic parameters of the underlying stochastic process. A formal
expansion of the current around the white-noise limit not always gives the
correct asymptotic behaviour. We improve the expansion for some simple but
representative cases.

\vspace*{1cm}

\vspace*{2cm}
\newpage
\topskip 0cm

\Section{Introduction}
In the last two years there has been a considerable interest
in the problem of noise induced transport. The motivation to
study such models has been initiated by Magnasco \cite{Magnasco}.
He showed that there are two necessary ingredients for noise
induced transport: A stochastic
force that is correlated in time and an environment
without inversion symmetry. In the one-dimensional case one
can have e.g. a periodic potential without inversion symmetry.
Such a potential is often called a ratchet-like potential.

Magnasco was mainly interested in the adiabatic limit, i.e.~
in the case where the fluctuations of the stochastic force are
slow. A few month later, the problem was investigated more carefully
by Doering et al.~\cite{Doering}. They studied the motion
of a one-dimensional particle in a saw-tooth potential and
in the case where the force is described by various stochastic
processes. For the symmetric dichotomous Markov process 
and at vanishing temperature they obtained an exact solution for
small correlation times. For the Ornstein--Uhlenbeck process
and for a class of processes called kangaroo processes, they
calculated the current to first order in the correlation
time. They also presented some
results from Monte-Carlo simulations of such a system,
again at zero temperature. 
Later, Mielke \cite{M1} developed a method that allows to
calculate the current for a large class of processes including the
ones discussed by Doering et al.~\cite{Doering}, again in the
case of a saw-tooth potential or more generally for a piecewise
linear potential. He recovered the results from \cite{Doering}
and found several other
cases where a current reversal occurs. One of these cases was
a process that consists of an even sum of dichotomous processes.
For such a process his results differ from the perturbative
result to first order in the correlation time \cite{Doering}.
The sign of the current was different, even for very small values
of the correlation time. One of the motivations of the present
work is this discrepancy. We will show that the current and
the stationary distribution of the coordinate of the moving particle
are non-analytic functions of the correlation time. To do this we
generalize the method developed in \cite{M1} so 
that it is applicable to zero temperature as well. 
This allows us to obtain exact numerical results
for the current and the stationary distribution, as well as a
correct asymptotic expansion for small correlation times.
This is also of general interest, since there are many approximative
methods to solve Langevin or Fokker--Planck equations in this
or other cases that agree with the usual perturbative expansion
for small correlation times (for a review we refer to \cite{Risken}).
Furthermore
we show that some of the properties of the solution
for small correlation times are relevant for larger correlation times
as well. Even if the solution is analytic for small
correlation times, it may become non-analytic for larger correlation times.
This always happens if the support of the stationary distribution
of the stochastic force is finite.

The paper is organized as follows: In the next section we
define the model and the stochastic processes we will be able
to deal with. In the subsequent section we show how the method 
in \cite{M1} can be generalized so that one can treat 
the zero temperature case as well.
Since a large part of the method is similar to the finite temperature case, 
we refer to \cite{M1} for details. In section 4 we present
some of the numerical results including a detailed discussion
of the current reversal for various noise processes. In section 5
we show how the asymptotic expansion for small correlation times
can be obtained and we give some explicit results for simple
noise processes. We also include a comparison between the
asymptotic expansion and the exact numerical results to show the
range of validity of the asymptotic expansion. Finally we give some
conclusions.

\Section{Definition of the model}
We consider the one-dimensional motion $x(t)$ of a particle in a dissipative
environment. The particle moves in a one-dimensional periodic potential
$V(x)$ with period $L$, 
and it is subject to a stochastic force $z(t)$ that is
correlated in time. In the over-damped regime, the motion of the
particle is described by a Langevin equation of the form. 
\beq \label{Lange1}
\frac{dx}{dt}=f(x)+\sqrt{2T}\xi(t)+z(t).
\eeq
We use units where the friction constant is unity. 
The first term on the right hand side is the force
$f(x)=-\frac{dV}{dx}$ due to the potential $V(x)$.
The second term describes thermal fluctuations, 
$\xi(t)$ is a white noise with zero mean and
$\average{\xi(t)\xi(t^{\prime})}=\delta(t-t^{\prime})$.
In the later part of the paper we will discuss the
zero temperature case, $T=0$. 
The additional noisy force $z(t)$ has zero mean,
$\average{z(t)}=0$.
We assume that it is described by a 
Markov process with an infinitesimal generator $M_z$. The probability
density $p(z,t)$ of this Markov process satisfies the 
Fokker-Planck equation
\beq \label{np1}
\frac{\partial p(z,t)}{\partial t}=M_z p(z,t).
\eeq
We discuss a class of Markov processes for which
the generator $M_z$ is described by its eigenvalues and
by certain properties of its right eigenfunctions,
namely
\bem
M_z\phi_n(z)=-\lambda_s\phi_n(z),\quad
z\phi_0(z)=\gamma_{0,1}\phi_{1}(z),
\eem
\beq \label{cop}
z\phi_n(z)=\gamma_{n,n+1}\phi_{n+1}(z)+\gamma_{n,n-1}\phi_{n-1}(z)
\quad n=1,2,\dots
\eeq
The eigenvalues $-\lambda_n$ obey 
\beq
\lambda_n \le \lambda_m \quad\mbox{if}\quad n<m,\quad 
\lambda_0=0
\eeq
$\phi_0(z)$ is the stationary distribution of $z$. 
Due to the recursion relations the eigenfunctions 
$\phi_n(z)$ can be written as $\phi_n(z)=g_n(z)\phi_0(z)$
where $g_n(z)$ are orthogonal polynomials with respect to the
weight function $\phi_0(z)$,
\beq
\int dz\,g_n(z)g_m(z)\phi_0(z)=\delta_{n,m}.
\eeq
This class of Markov processes is very general. It contains
many processes that occur in typical situations such as
the Ornstein--Uhlenbeck process, the dichotomous process,
sums of dichotomous processes, and kangaroo processes.
The correlation time $\tau$ of the process is related to
the smallest non-vanishing eigenvalue $\lambda_1$ of the noise process
via $\tau=\lambda_1^{-1}$.

The joint probability density $\rho(x,z,t)$
for the two stochastic variables $x(t)$ and $z(t)$ obeys a 
Fokker-Planck equation of the form
\beq \label{fp1}
\frac{\partial \rho(x,z,t)}{\partial t}=
-\frac{\partial}{\partial x}
\left(f(x)+z\right)\rho(x,z,t)
+M_z \rho(x,z,t)
\eeq
We discuss only stationary properties of this equation,
therefore the left hand side is put to zero and 
(\ref{fp1}) becomes an equation for the stationary
probability density $\rho(x,z)$. Due to the periodicity of
$V(x)$ we assume $\rho(x,z)$ to be periodic in $x$
with period $L$. To solve the stationary Fokker-Planck equation it
is useful to expand $\rho(x,z)$ in terms of the right eigenfunctions of
$M_z$.
\beq
\rho(x,z)=p_0(x)\phi_0(z)
+\sum_{n=1}^{\infty}(-1)^n \phi_n(z)p_n^{\prime}(x).
\eeq
This yields the recursion relations
\beq \label{recur1}
J=f(x)p_0(x)-\gamma_{0,1}p_1^{\prime}(x)
\eeq
\beq \label{recur2}
\gamma_{0,1}p_0(x)=\lambda_1 p_1(x)+f(x)p_1^{\prime}(x)
-\gamma_{1,2}p_2^{\prime}(x)
\eeq
\beq \label{recur3}
\gamma_{n-1,n}p_{n-1}^{\prime}(x)=\lambda_n p_n(x)+f(x)p_n^{\prime}(x)
-\gamma_{n,n+1}p_{n+1}^{\prime}(x),\quad n>1.
\eeq
for the functions $p_n(x)$.
As additional conditions we have the normalization of $p_0(x)$ and the
periodicity of $p_n(x)$,
\beq
\int_0^Lp_0(x)dx=1, \quad p_n(x)=p_n(x+L).
\eeq
$p_0(x)$ is the stationary distribution of $x(t)$. It is one of
the quantities we are interested in. A second and more important
quantity is the integration constant $J$ in (\ref{recur1}),
it is the stationary current. If the potential $V(x)$ has inversion
symmetry, the current vanishes. In the general situation, 
where $V(x)$ has no inversion symmetry, one obtains generically
a non-vanishing current.

To solve the recursion relations (\ref{recur1}--\ref{recur3})
for a general Markov process $M_z$ approximations are necessary.
One has to truncate the recursion at some large value $N$, i.e.~
one has to put $\gamma_{N,N+1}=0$. For some
processes, e.g. for a sum of dichotomous processes, one has 
$\gamma_{N,N+1}=0$ for some finite value of $N$ and the approximation
is not necessary. In principle it is then
possible to solve the recursion relations numerically. 
Here, we restrict ourselves to a class of piecewise linear
potentials. This is a standard assumption \cite{Doering}, such a 
potential is often called a ratchet-like potential. If
the potential is piecewise linear, the force $f(x)$ is
piecewise constant. The differential equations (\ref{recur1}--\ref{recur3})
are equations with constant coefficients that can be solved
analytically. The remaining algebraic problem is to solve some continuity
conditions for the functions $p_n(x)$ at the points where the
force jumps from one constant value to another. In that way it is possible
to express the stationary current as a ratio of two determinants.
It can thus be calculated easily numerically for various parameters
of the system or even analytically for small values of $N$. 
The general procedure has been described for
finite temperatures in detail in \cite{M1}. 
One can obtain 
very accurate results in the whole parameter regime and for
various noise processes. Thereby one observes that
several analytical approximation schemes do not work in different
regimes of the parameter space. For instance, the standard expansion
for small correlation times $\tau$ breaks down. The physical reason is
that for sufficiently small values of $z$ the particle cannot escape 
one of the minima of the potential. At zero temperature, this is
only possible if $\abs{z}$ is larger than the largest value of $\abs{f(x)}$. 
Since the process $z(t)$ has a finite correlation time, 
$z$ remains for a while in the region where $\abs{z}<\abs{f(x)}$
and the particle cannot move apart. For a general potential
$V(x)$ this yields divergencies in the stationary distribution
$p_0(x)$, in the case of a piecewise linear potential,
the stationary distribution contains contributions of the type
$w_i\delta(x-x_i)$, where $x_i$ are the positions of the minima
of $V(x)$ and $w_i$ are some weights. Such an effect occurs 
always if there is a finite probability to have a stochastic force $z(t)$, 
that fulfils the condition $f(x_i-0_+)\le z \le f(x_i+0_+)$.
 It occurs especially in the
case where $\gamma_{N,N+1}=0$ for some finite $N$. Mathematically
it turns out that the stationary distribution $p_0(x)$ and the current
$J$ are nonanalytic functions of $\tau$ in some regions of the
parameter space. This is the reason why the $\tau$--expansion fails
for some noise processes. 

In the next section we describe what modifications to 
the method in \cite{M1} are necessary so that $T=0$ can be treated as well.
The main goal is to obtain correct zero temperature
results for the stationary behaviour of the model and to
derive a correct asymptotic expansion for small $\tau$.

\Section{The method at zero temperature }
In the rest of the paper we restrict ourselves to the simplest
nontrivial case for the force $f(x)$; we assume that $f(x)$ takes to
different values. To be precise, we let 
\beqar f(x)=f_1
&\quad\mbox{if}\quad& 0\le x<L_1, \ret f(x)=f_2 &
\quad\mbox{if}\quad&
L_1\le x<L.  
\eeqar 
We let $L_2=L-L_1$. Due to the periodicity of
$V(x)$ we have $f_1L_1+f_2L_2=0$. We assume that $V(x)$ has a minimum
at $x=0$, which means $f_1<0$, $f_2>0$. Due to the discussion above,
$p_0(x)$ has the form 
\beq 
p_0(x)=\tilde{p}_0(x)+W_{0}\,\delta(x).
\eeq 
$\tilde{p}_0(x)$ contains no further $\delta$--contributions. In the same way
we obtain for the functions $p^\prime_n(x)$ 
\beq
p^\prime_n(x)=\tilde{p}^\prime_n(x)+W^\prime_{n}\,\delta(x).  
\eeq 
The coefficients $W_0$ and $W^\prime_{n}$ can be related to each other
using the differential equations \ref{recur1}-\ref{recur3}. One obtains 
\beqar \label{WBez}
W^{\prime}_{2k-1}&=&0 \\
W^{\prime}_{2k}&=&\left(-1\right)^k\,
\frac{\gamma_{2k-2,2k-1}}{\gamma_{2k-1,2k}}
\cdots\frac{\gamma_{0,1}}{\gamma_{1,2}}\,W_0\label{WBez2} 
\eeqar 
For $\tilde{p}_0(x)$ and $\tilde{p}^\prime_n(x)$ we make the ansatz
\beqar\label{fund} \tilde{p}_{0,i}(x)&=& \sum_r
c_{r,i}a_{0,i}^{(r)}\alpha_i^{(r)}e^{\scriptstyle(\alpha_i^{(r)}x)}
+b_{0,i}\\
\tilde{p}_{1,i}(x)&=&
\sum_r c_{r,i}a_{1,i}^{(r)}e^{\scriptstyle(\alpha_i^{(r)}x)}+b_{1,i}\\
\tilde{p}_{n,i}(x)&=& \sum_r
c_{r,i}a_{n,i}^{(r)}e^{\scriptstyle(\alpha_i^{(r)}x)}+b_{n,i} 
\eeqar
The index $i$ takes the two values $i=1,\,2$, according to the two
regions where $f(x)=f_i$. Inserting this ansatz in the differential
equations, we obtain a generalized eigenvalue problem to determine the
coefficients $\alpha_i^{(r)}$, $a_{n,i}^{(r)}$, and $b_{n,i}^{(r)}$.
This is the usual procedure for a system of coupled linear
differential equations with constant coefficients and has been
described for the present case in detail in \cite{M1}.  We obtain using
a vector notation 
\beq\label{b} {\bf \vec{b}}_i=
\left(\!\begin{array}{c}
    b_{0,i}\\b_{1,i}\\b_{2,i}\\\vdots\end{array}\!\right)=
\left(\!\begin{array}{c}
    \frac{J}{f_i}\\\frac{J\gamma_{0,1}}{\lambda_1f_i}
    \\0\\
    \vdots\end{array}\!\right)\quad .  
\eeq 
The generalized eigenvalue
problem for $\alpha_i^{(r)}$ and $a_{n,i}^{(r)}$ is 
\beq \label{evp}
{\bf A}_i\,\vec{\bf a}_i=\alpha_i\,{\bf B}_i\,\vec{{\bf a}}_i\quad .
\eeq 
where 
\beq 
{\bf A}_i=\left(\begin{array}{cccc}
    0&0&0&\ldots\\
    0&\lambda_1&0&\ldots\\
    0&0&\lambda_2&\ddots\\
    \vdots&\ddots&\ddots&\ddots\end{array}\right)\quad ,\quad 
{\bf B}_i=\left(\begin{array}{cccc}
    -f_i&\gamma_{0,1}&0&\ldots\\
    \gamma_{0,1}&-f_i&\gamma_{1,2}&\ldots\\
    0&\gamma_{1,2}&-f_i&\ddots\\
    \vdots&\ddots&\ddots&\ddots\end{array}\right)\quad .  
\eeq
$\alpha_i^{(0)}$ is zero, the other eigenvalues can be determined
either be explicitely solving the eigenvalue problem or by using
certain continued fractions as described in \cite{M1}. To finally solve
the problem, one has to determine the coefficients $c_{r,i}$, the
weight $W_0$ of the $\delta$--contribution in $p_0(x)$, and the current
$J$. These are $2N+2$ unknown variables. The continuity of $p_n(x)$ at
$x=L_1$ and certain jump conditions of $p_n(x)$  at $x=0$ involving 
the coefficients $W^\prime_{n}$ for $n\ge 1$ yield $2N$ equations, 
in addition we have the normalization of $p_0(x)$. 
The last condition can be obtained
from the continuity of the current density at a fixed value of $z$.
For finite $N$ we can write $\rho(x,z)$ in the form 
\beq
\rho(x,z)=\sum_{k=0}^N\, P^{(k)}(x)\,\delta(z-z_k)\quad .  
\eeq 
for a
process $z(t)$ that takes $N+1$ different values $z_k$.  $P^{(k)}(x)$
can be related to $p_0(x)$ and $p^\prime_n(x)$, for instance 
\beq
\sum_{k=0}^N\,P^{(k)}(x)=p_0(x).  
\eeq 
For each $P^{(k)}(x)$ we have a corresponding current 
\beq 
J^{k}(x):=P^{k}(x)\,f^{k}(x).  
\eeq 
We introduced $f^{k}(x):=f_i+z_k$ if $x$ lies in the region where
$f(x)=f_i$. The current $J^{k}(x)$ has to be continuous at $x=L_1$.
If $\abs{z_k}<\min_i(\abs{f_i})$ and $\mbox{sign}(z_k)=-\mbox{sign}(f_i)$
 this yields $J^{k}(L_1)=0$.
With this equation we have found the last condition we need to determine the
$2N+2$ unknown variables. In this way the calculation of the current
$J$ and the stationary distribution $p_0(x)$ has been reduced to the
purely algebraic problem of solving $2N+2$ linear equations for $2N+2$
variables. As in \cite{M1}, the current $J$ can be expressed in closed
form using determinants.
\Section{Results}
The results presented in this section have been obtained
by solving the algebraic problem mentioned above numerically.
We present results for two classes of processes, namely for
a sum of dichotomous processes and for some kangaroo processes.

\subsection{Sums of dichotomous processes}
A dichotomous process $z(t)$ is a process where $z(t)$ takes
two values, $\pm z_0$. Summing up $N$ such processes, we
obtain a Markov process with a stationary distribution
of the form 
\beq\label{zstat}
\phi_0(z)=\frac{1}{2^N}\left(\sum\limits_{i=0}^N\,{N\choose i}
\delta\left(z-(N-2i)\,z_0\right)\right).
\eeq
The parameters that characterize the process are
\beq
\lambda_n=-n/\tau, \quad n=0,\dots, N,
\eeq
\beq
\gamma_{n,n+1}=\sqrt{{(n+1)(N-n)}}z_0, \quad n=0,\dots, N.
\eeq
In order to have a value for $\average{z^2}=\gamma_{0,1}^2$ that
is independent of $N$, we choose $z_0=\gamma/\sqrt{N}$ where
$\gamma=\sqrt{D/\tau}$. $D$ is the noise strength of the process.
In the limit $N\rightarrow\infty$, the sum of $N$ dichotomous
processes yields the Ornstein--Uhlenbeck process \cite{shi}. 

The eigenvalue problem (\ref{evp}) can be solved explicitely for a sum of
dichotomous processes. This has been shown in \cite{M1}. The eigenvalues
$\alpha_i^{(k)}$ are 
\beq\label{alpha1} 
\alpha^{(1)}_i=\frac{N\lambda f_i}{N\gamma^2-f_i^2} 
\eeq 
\beq\label{alpha2}
\alpha^{(2k),(2k+1)}_i=\frac{N\lambda f_i}{2}\frac{N\pm (N-2k)
\sqrt{1+\frac{4k(N-k)\gamma^2}{Nf_i^2}}}{(N-2k)^2\,\gamma^2-Nf_i^2}\ 
,\ \mbox{for $2k+1\leq N$}
\eeq 
and in addition 
\beq\label{alpha3}
\alpha^{(N)}_i=-\frac{N\lambda}{2f_i} 
\eeq 
for even $N$. The
coefficients $J$, $W_0$, and $c_{r,i}$ can be determined numerically
as described above. Let us first discuss the results for small
correlation times $\tau$. Since $\gamma^2=D/\tau$, $\lambda=1/\tau$,
the eigenvalues $\alpha_i^{(k)}$ have singularities
$\propto\tau^{-1/2}$ and $\propto\tau^{-1}$. The second type of
singularity is present only if $N$ is even. Furthermore, for small
$\tau$ and $N$ odd, $\abs{z(t)}$ is always larger than $\abs{f(x)}$
and we expect that $W_0=0$, whereas for even $N$, $z(t)$ may be zero
and therefore we expect $W_0>0$. The behaviour should be very
different depending on whether $N$ is even or odd.  First we will show
some results for the current.  In Fig. 1a the current 
is plotted as a function of $\tau$ for $N$
odd. The current is always positive.  The corresponding result for
even $N$ is shown in Fig. 1b. The current is
negative for small $\tau$ and changes the sign when $\tau$ becomes
larger. This has already been observed for very low temperatures in
\cite{M1}. We will come back to this point in the next section, where we
discuss an expansion for small $\tau$. Let us mention that a negative
current for even $N$ and small $\tau$ occurs only if the noise
strength $D$ is sufficiently large. For larger values of $\tau$, the
current is shown for some processes in Fig. 2. The
current has cusp-like maxima that occur near such values of $\tau$,
for which one of the discrete values of $\abs{z(t)}$ becomes
smaller than one of the discrete values of $\abs{f(x)}$. The exact
position of these values of $\tau$ is indicated in the figure. It is
interesting to see that the maxima don't lie exactly at this value
of $\tau$ but slightly below. With increasing number of maxima, 
i.e.~for higher values of $\tau$ this effect increases as well. 
Up to now we have no physical interpretation for it. 

Whenever $\tau$ becomes larger than one
of these values of $\tau$, the weight $W_0$ of the $\delta$--peak in
the stationary distribution $p_0(x)$ jumps discontinuously to a higher
value. This is shown in Fig. 3 for the same parameter
as used in Fig. 2. The weight $W_0$ shows the expected behaviour and
is unity for large $\tau$.  The coefficients $c_{r,i}$ can be
calculated as well, they enter in the expression for 
the stationary distribution $p_0(x)$. Typical results are shown for
various processes in Fig. 4. $\tau$ is very small, so
that a $\delta$--contribution (which is not shown) occurs only for
even $N$. One observes that for both cases, even or odd values of $N$,
the stationary distribution $p_0(x)$ has discontinuities. 
We will come back to that point later. 
The minimum at $x=0.8$ is located at the maximum of
the potential. It is strong for $N=2$ and becomes weaker when $N$
is even and increases. The opposite behaviour is found for odd $N$,
here the minimum is only weak for $N=1$ and becomes more and more
pronounced for larger $N$. Similarly, the maximum at $x=0$ increases
with increasing $N$ if $N$ is odd. 

\subsection{Kangaroo processes}
Kangaroo processes are processes with $\lambda_n=-1/\tau$ for all
$n>0$. They can be completely characterized by the stationary
distribution $p_{st}(z)$. In this section we discuss results for
kangaroo processes with a stationary distribution given in (\ref{zstat}). 
 According to the notations there, we call these processes $K(N)$.
 This allows a direct comparison with the results for the sums of
dichotomous processes. The only difference between the kangaroo
processes and the sums of dichotomous processes is that for the latter
case, only jumps of a magnitude $\pm z_0$ in $z(t)$ occur, whereas a
kangaroo process has no restriction of the jumps of $z(t)$.  Since the
stationary distributions are the same, the parameters $\gamma_{n,n+1}$
are the same. But unfortunately, one cannot derive an analytic
expression for the eigenvalues $\alpha_i^{(r)}$ for $N>4$. They have
to be determined numerically, which is easily done.

We show some of the numerical results for the current as a function of
the correlation time $\tau$ in Fig.~5a for $N$ odd and Fig.~5b for $N$
even. The inset in Fig.~5a shows that
the current is negative for very small $\tau$. This is in contrast to
the sums of dichotomous processes.  Furthermore, the region of
negative current becomes larger, when $D$ becomes smaller. For even
$N$ the current reversal is observed as for the sum of dichotomous
processes, but the region of negative current is larger. Furthermore
the current is always negative for small $\tau$ and $N$ even, whereas
for the sum of dichotomous processes a negative current occurs only if
the noise strength is large. Thus, the behaviour of the current as a
function of $\tau$ and $D$ differs significantly from that one of the sums of
dichotomous processes, although the stationary distribution is the
same in both cases.  Let us mention that the mechanism that produces a
negative current for small $\tau$ and $N$ odd is not clear. Doering et
al.~\cite{Doering} proposed that a negative current occurs due to a
parameter called flatness, which is a property of the stationary
distribution. Their argument cannot be applied in the present case,
since the stationary distributions and therefore the flatness is the
same for the kangaroo processes and the sums of dichotomous
processes, whereas the sign of the current for small $\tau$ is different.

\Section{An expansion for small $\tau$}
The usual formal expansion for small $\tau$ has already been
discussed by Doering et al \cite{Doering}. It can be obtained
using a standard perturbational treatment \cite{Kato}
to solve the stationary Fokker--Planck equation.
The unperturbed part is $M_z$, which is of order
$\tau^{-1}$. The perturbation contains a term proportional to
$z$, it is thus $\propto\tau^{-1/2}$. But since it contains only
off-diagonal matrix elements in the basis in which $M_z$ is
diagonal (there are no parameters $\gamma_{n,n}$), one
obtains finally an expansion in powers of $\tau$ for the
stationary distribution and for the current $J$. Since the
current vanishes in the white noise limit, the first term in
the expansion for $J$ vanishes and $J\propto\tau$. 
An alternative way to obtain the same expansion is to use
a formal operator continued fraction that can be obtained
from the recursion relations (\ref{recur1}-\ref{recur3}). 
This has been shown
in the appendix of \cite{M1}, where a general expression for the first
order has been given and higher orders can be obtained straight forward. 
In our special case of a saw--tooth potential one obtains   
\beq
J=\tau J_1+\tau^2J_2 +\dots
\eeq
with
\beq\label{jform1}
J_1= \left(1-\frac{\displaystyle \lambda_1}{\displaystyle \lambda_2}\,
\frac{\displaystyle \gamma_{1,2}^2}{\gamma_{0,1}^2}\right)\,
\frac{\displaystyle f_1^2-f_2^2}{D^3\,
\left(\frac{\displaystyle 1}{\displaystyle f_1}-
\frac{\displaystyle 1}{\displaystyle f_2}\right)^2
\left(e^{\textstyle \frac{1}{D}}-1\right)
\left(e^{\textstyle \frac{-1}{D}}-1\right)} 
\eeq
\beqar\label{jform2}
J_2&=&\left(1-\frac{\displaystyle \lambda_1}
{\displaystyle \lambda_2}\,
\frac{\displaystyle \gamma_{1,2}^2}{\gamma_{0,1}^2}\right)\:
\frac{\displaystyle f_1^2-f_2^2}{\displaystyle N_{inh}^{(1)}\,D^2\,
\left(\frac{1}{f_1}-\frac{1}{f_2}\right)\:
\left(e^{\frac{1}{D}}-1\right)}
\ret
&&+\left(2\,\frac{\gamma_{1,2}^2}{\gamma_{0,1}^2}\,
\frac{\lambda_1}{\lambda_2}+\frac{\lambda_1^2}{\lambda_2^2}\,
\frac{\gamma_{1,2}^2}{\gamma_{0,1}^2}-\frac{\lambda_1^3}
{\lambda_2^2\lambda_3}\,\frac{\gamma_{2,3}^2\gamma_{1,2}^2}
{\gamma_{0,1}^4}-\frac{\lambda_1^2}{\lambda_2^2}\,
\frac{\gamma_{1,2}^4}{\gamma_{0,1}^4}-1\right)
\ret
&&\ast\frac{\displaystyle f_1^4-f_2^4}{D^4\,
\left(\frac{\displaystyle 1}{\displaystyle f_1}-
\frac{\displaystyle 1}{\displaystyle f_2}\right)^2
\left(e^{\textstyle \frac{1}{D}}-1\right)
\left(e^{\textstyle \frac{-1}{D}}-1\right)}.
\eeqar
This result generalizes the result given by
Doering et al.~\cite{Doering}. It can be
compared with the exact results of the sums of 
dichotomous processes and for the kangaroo processes
shown in the last section. One observes that the formal
expansion is correct for a sum of an odd number of
dichotomous processes, if the correlation time $\tau$ is sufficiently small.
For a sum of an even number of dichotomous processes the
result is not even qualitatively correct, the current has
the wrong sign. Similar results hold for the
kangaroo processes. If $N$ is odd, the expansion is
correct for small $\tau$, whereas it is wrong for
even $N$. The reason is that the stationary distribution
$p_0(x)$ contains a contribution $W_0\delta(x)$ for even $N$
and arbitrary small $\tau$, but not for 
odd $N$. This leads to non-analytic contributions to $p_0(x)$ 
and $J$ as functions of $\tau$ if $N$ is even, which are not taken
into account in the perturbative $\tau$--expansion. This expansion
is well defined only if $p_0(x)$ is a differentiable function of $x$,
which is true for odd $N$, but not for even $N$. 

A second possibility to obtain an asymptotic $\tau$--expansion
for $J$ and $p_0(x)$ is to start from the set of linear
equations that determine the variables $J$, $W_0$,
and $c_{r,i}$. The coefficients in this set of linear
equations contain singularities of the type
\[
e^{\displaystyle+\frac{\displaystyle c}
{\displaystyle \sqrt{\tau}}}\quad\mbox{(type 1)} 
\]
\[
e^{\displaystyle -\frac{\displaystyle c}
{\displaystyle \sqrt{\tau}}}\quad\mbox{(type 2)}
\]
\[e^{\displaystyle -\frac{\displaystyle c}{\displaystyle \tau}}
\quad\mbox{(type 3)}
\]
where $c$ is some constant.
These singularities occur due to
the behaviour of the eigenvalues $\alpha_i^{(k)}$ for small
$\tau$, see e.g.~(\ref{alpha1}-\ref{alpha3}). 
For small $\tau$ it is possible to
divide those equation, which contain singularities of
type 1 by $\exp(+c/\sqrt{\tau})$. We can then neglect all the terms
that vanish faster than any power of $\sqrt{\tau}$. For small
$N$ it is possible to solve the remaining set of linear
equations analytically. For a single dichotomous process
this yields
\beq
J^{(1)}(\tau)=J_1^{(1)}\,\tau+J_2^{(1)}\,\tau^2+{\cal O}(\tau^3)
\eeq
where
\beqar
 J_1^{(1)}&=& 
\frac{\displaystyle f_1^2-f_2^2}{D^3\,
\left(\frac{\displaystyle 1}{\displaystyle f_1}-
\frac{\displaystyle 1}{\displaystyle f_2}\right)^2
\left(e^{\textstyle \frac{1}{D}}-1\right)
\left(e^{\textstyle \frac{-1}{D}}-1\right)}\\  
 J_2^{(1)}&=&\frac{f_2^4+f_1^4}{\displaystyle D^6\,
\left(\frac{1}{f_1}-\frac{1}{f_2}\right)^2
\left(e^{\frac{1}{D}}-1\right)^2\,\left(e^{-\frac{1}{D}}-1\right)^2}
\ret
&&+\frac{1}{2}\,\frac{\left(f_2^4+f_1^4\right)\:
\left(1-e^{\frac{2}{D}}\right)}{\displaystyle D^5\,
\left(\frac{1}{f_1}-\frac{1}{f_2}\right)^2
\left(e^{\frac{1}{D}}-1\right)^4}
\ret
&&-\frac{f_1^4-f_2^4}{\displaystyle D^4
\left(\frac{1}{f_1}-\frac{1}{f_2}\right)^2
\left(e^{\frac{1}{D}}-1\right)\,\left(e^{-\frac{1}{D}}-1\right)}.
\eeqar
As expected, this result agrees with the formal $\tau$--expansion
described above, which was correct for a single dichotomous process.
For a sum of two dichotomous processes we obtain
\beq\label{dich2reihe} 
J^{(2)}=\frac{1}{2}J_1^{(1)}(1-D)\tau+{\cal O}(z^3).
\eeq
This expression differs from the formal $\tau$--expansion, but
it agrees well with the exact numerical results for small $\tau$. 
The current is negative for small $\tau$ and $D>1$. 
For $N=3$ one would again expect a result that agrees with
the formal $\tau$--expansion, but this is not true. We obtain
\beqar
 J^{(3)}&=&\frac{1}{3}J_1^{(1)}\tau+
J_{3/2}^{(3)}\tau^{\frac{3}{2}}+{\cal O}(z^4)
\label{dich3reihe}
\\
 J_{3/2}^{(3)}&=&-\frac{1}{2\sqrt{6}}\,D^{-\frac{5}{2}}
\frac{f_1^2f_2^2\,(f_1-f_2)}{\left(e^{\frac{1}{D}}-1\right)
\left(e^{\frac{-1}{D}}-1\right)}.
\eeqar
The additional term $\propto\tau^{\frac{3}{2}}$ was not present in the
formal $\tau$--expansion. It is responsible for the very small
region of validity of the formal $\tau$--expansion for $N=3$. Similar
terms occur for any odd $N$. The reason is that the stationary
distribution $p_0(x)$ is not a differentiable function of $x$.
It has discontinuities at the extrema of the potential, see Fig. 4. Therefore
the perturbative $\tau$--expansion is not well defined.

Finally one can look at the kangaroo process $K(2)$, for which the
formal $\tau$--expansion was incorrect too. We obtain
\beqar
J^{K(2)}&=&J_1^{K(2)}\,\tau+{\cal O}(z^3)\\
J_1^{K(2)}&=&-D\,J^{(1)}_1,
\eeqar
which is in agreement with the exact numerical results.
A comparison between the numerical results and the asymptotic
expansions is shown in Fig. 6.

\Section{Conclusions}
We discussed in detail the behaviour of the static properties of 
one-dimensional models for noise induced transport at zero temperature.
For our discussion we used a sawtooth potential, but the method
and the results can be generalized easily to any piecewise linear potential.
If the support of the stationary distribution of the stochastic force
$z(t)$ is finite, one observes non-analytic contributions in the
induced current and in the stationary distribution of the coordinate
of the particle. In a piecewise linear potential, the stationary 
distribution of the coordinate $p_0(x)$ contains contributions of the form
$w_i\delta(x-x_i)$. Such contributions occur for sufficiently large values
of $\tau$, for some processes even for all $\tau>0$. If the 
$\phi_0(z)$ and the force $f(x)=-dV/dx$ are continuous,
one can not expect a 
$\delta$--contribution to $p(x)$. But even then one expects singularities
in $p(x)$. Let $f_{\rm min}$ and $f_{\rm max}$ be the minimum
and the maximum of $f(x)$, and let us assume without loss of generality
that $\abs{f_{\rm min}}<\abs{f_{\rm max}}$. 
Let us now take a stochastic force 
$z(t)$ that takes only a finite, discrete set of values $z_i$. If
$0<z_i<-f_{\rm min}$, the particle moves to the right
until $z_i=f(x)$. At these special values of $x$ singularities in
$p(x)$ may occur, depending on the behaviour of $f(x)$ near such 
a point. A similar situation occurs if $0>z_i>-f_{\rm max}$. 
In the more general case where the stochastic force can take all
values within an interval $[-z_0,z_0]$, similar singularities
may occur, again depending on the form of $f(x)$. Therefore
the results we presented are relevant for a large class of situations.
Let us mention that one should not expect such singularities in
the case of a Gaussian noise process, since in that case the
stochastic force may be arbitrarily large. 

A consequence of the non-analytic behaviour of $p(x)$ is that
the usual perturbative $\tau$-expansion breaks down. In the
$n$-th order of this expansion a derivative of order $(n+2)$
of $p(x)$ occurs. If this derivative doesn't exist, it is clear
that the $\tau$-expansion is not defined. In contrary, the
asymptotic $\tau$-expansion we derived for some special cases is
always well defined.

\pagebreak


\pagebreak \thispagestyle{empty}
\setcounter{section}{1}

\noindent
\subsection*{Figure captions}

\renewcommand{\labelenumi}{{Fig.} \arabic{enumi}.}
\begin{enumerate}
\item The current $J$ as a function of the correlation
 time $\tau$ for a sum of $N$ dichotomous processes. 
a) $N$ odd and  
$D=10\,\Delta V$ and $L_1/L_2=4$.\\
 b) The same as in a) for N even.
\item  The stationary current for the sum 
of $N=1,2,3$ dichotomous processes as a function $\tau$ on 
a larger range of $\tau$. The tags $\tau_{Nk}$ indicate the
 correlation times, for which is $k\,z_0=f_2$. The curves
for $N=1,2,3$ have maxima near $\tau_{1k}, \tau_{2k},\tau_{3k}$.
The parameters are the same as in Fig.~1. 
\item The weight of the $\delta$ distribution as a function 
of $\tau$ for the sum of $N$ dichotomous  processes for the 
same parameters as in Fig.~1.
\item The stationary probability distribution of the position 
coordinate in a normed period and with $L_1/L_2=4$ for the sum 
of $N$ dichotomous processes.  $N=1$ (solid line), $N=2$ (dashed line)
$N=3$ (short dashed line) and $N=4$ (long dashed line). The 
$\delta$--contribution is not shown.
\item a) The current $J$ as a function of the correlation time $\tau$ for the kangaroo processes $K(N)$, $N$, odd with the same parameters as in Fig.~1. The inset shows the same plot on a finer scale.\\
b) The same as in a) for N even.
\item Comparison of the linear approximation (solid lines) of the current
as a function of the correlation time with the exact results (dashed lines) for the sum of $N$ dichotomous processes and the kangaroo process $K(2)$.   
\end{enumerate}

\pagebreak \thispagestyle{empty}



\begin{figure}[h]
\leavevmode
\centering
\setlength{\unitlength}{1cm}
\begin{picture}(15,12)
\put(6.5,12){\mbox{\Large\bf Fig.~1.a)}} 
\put(-0.5,10.5){\mbox{\large\boldmath $J/10^{-2}\Delta V$}}
\put(6,2){\mbox{\large\boldmath$\tau/10^{-2}\Delta V^{-1}$}}
\put(0,10.5){\mbox{
\epsfig{file=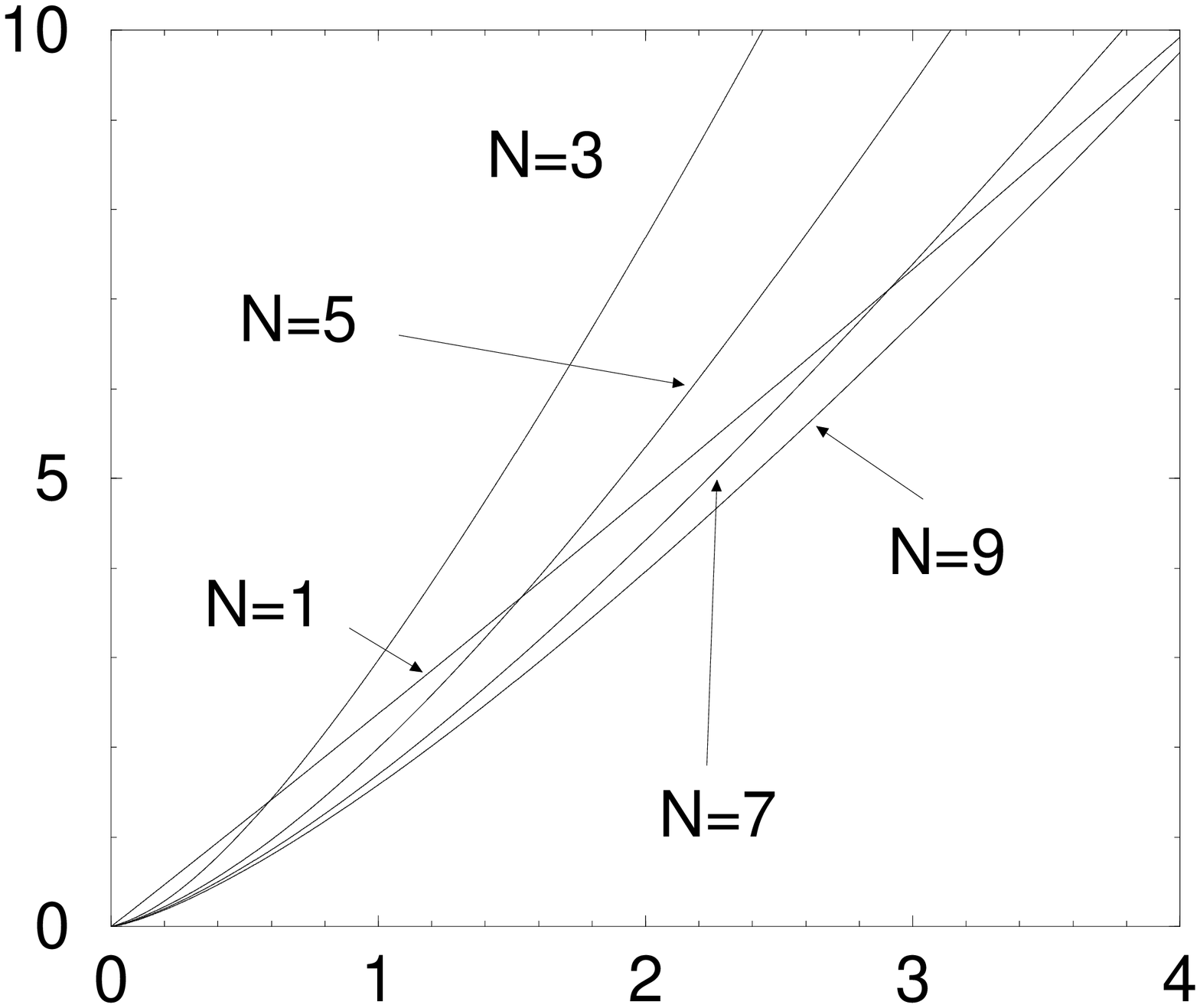,height=12.5cm,width=8cm,angle=270}}}
\end{picture}
\begin{picture}(15,10)
\put(6.5,10){\mbox{\Large\bf Fig.~1.b)}}
\put(-0.5,8.5){\mbox{\large\boldmath $J/\Delta V$}}
\put(6,0){\mbox{\large\boldmath$\tau/\Delta V^{-1}$}}
\put(0,8.5){\makebox{
\epsfig{file=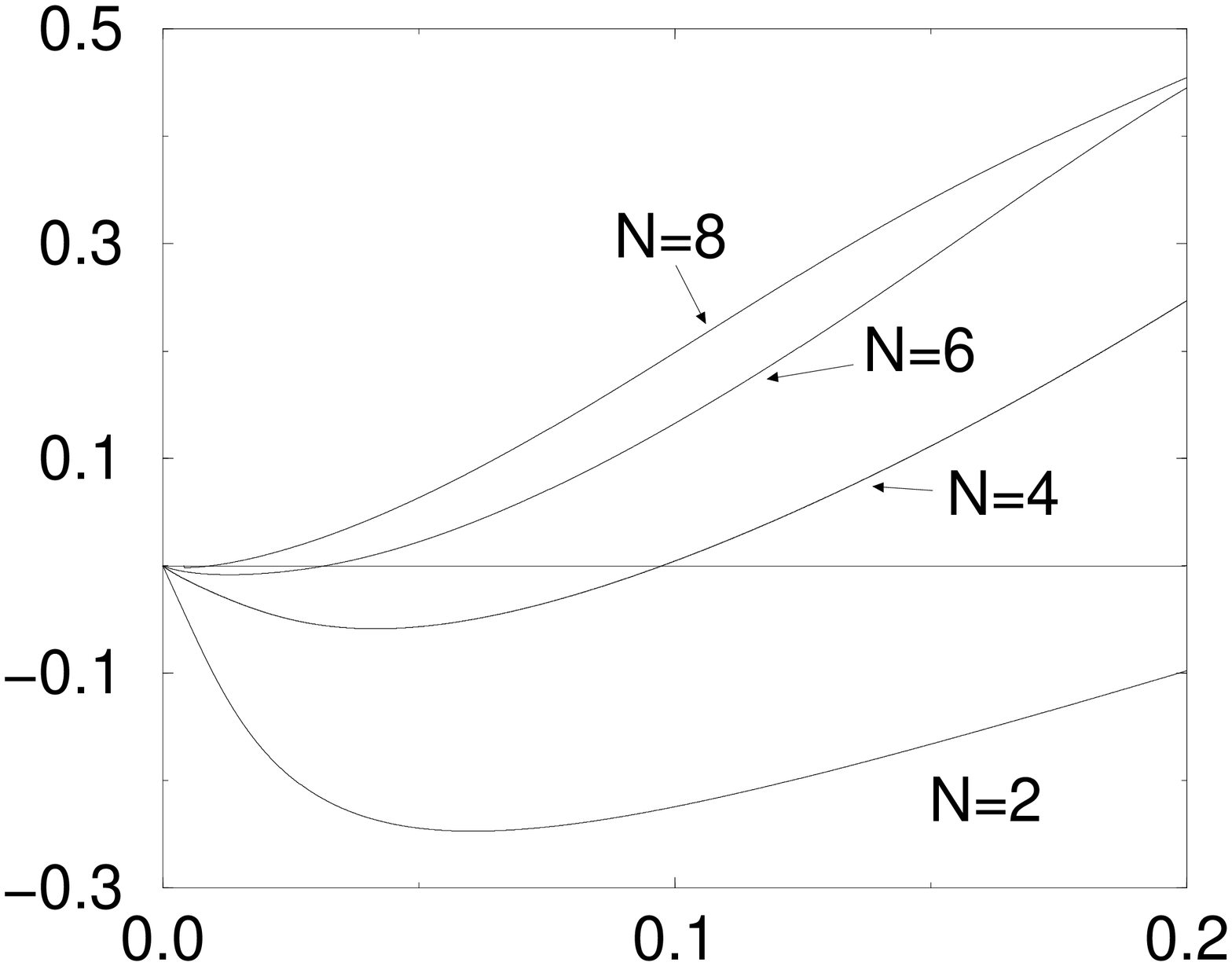,height=12.5cm,width=8cm,angle=270}}}\end{picture}\\
\end{figure}

\pagebreak


\begin{figure}[h]
\leavevmode
\centering
\setlength{\unitlength}{1cm}
\begin{picture}(15,15)
\put(6.5,15){\mbox{\Large\bf Fig.~2}} 
\put(0.5,7.5){\makebox{\large\boldmath $J/\Delta V$}}
\put(6.5,-0.2){\makebox{\large\boldmath $\tau/\Delta V^{-1}$}}
\put(0,14){\mbox{\epsfig{file=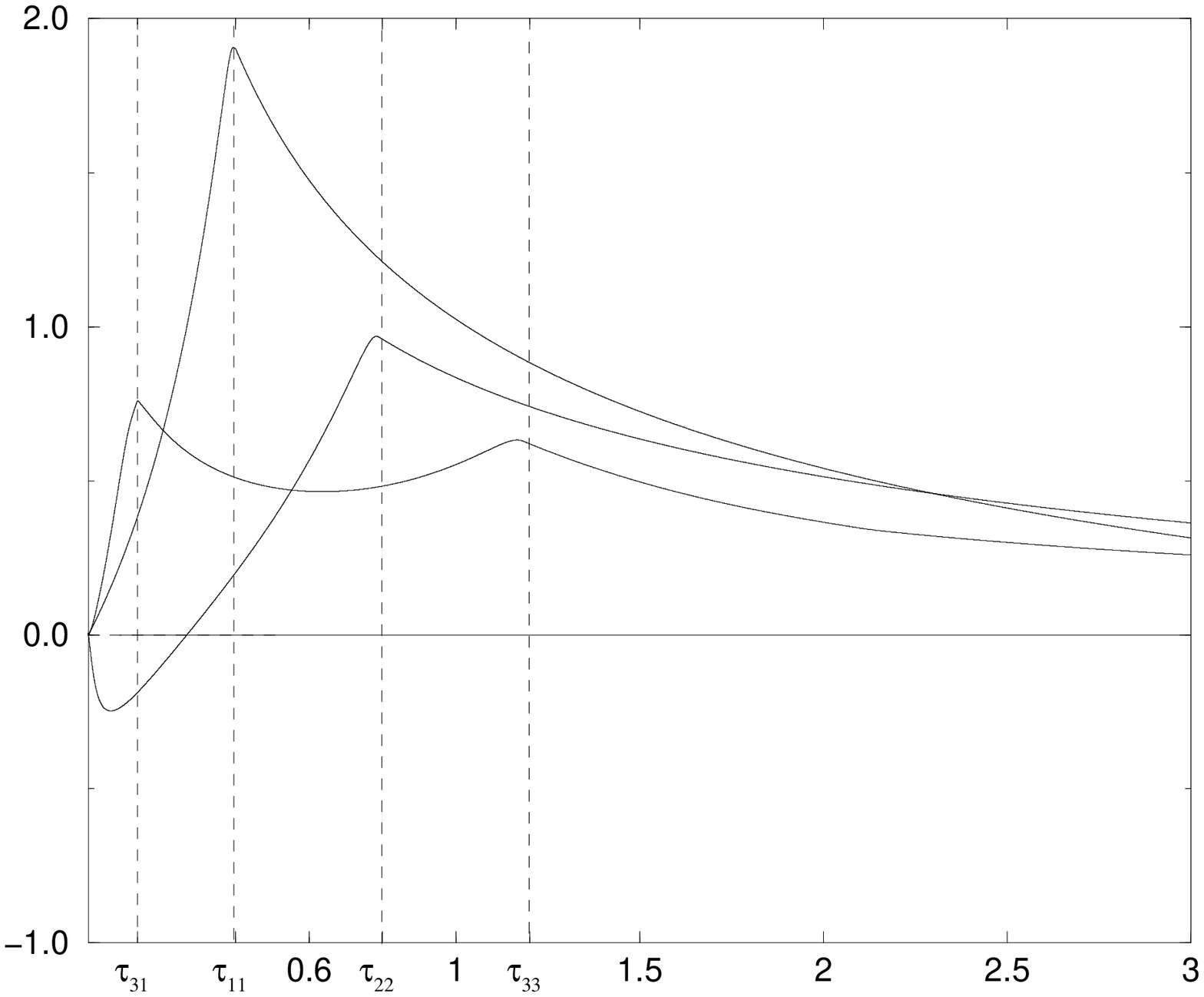,height=15cm,width=14cm,angle=270}}}
\end{picture}
\end{figure}

\clearpage


\begin{figure}[h]
\leavevmode
\centering
\setlength{\unitlength}{1cm}
\begin{picture}(15,15)
\put(6.5,15){\mbox{\Large\bf Fig.~3}} 
\put(0.5,7.5){\makebox{\large\boldmath $W_0$}}
\put(6.5,-0.2){\makebox{\large\boldmath $\tau/\Delta V^{-1}$}}
\put(0,14){\makebox{
\epsfig{file=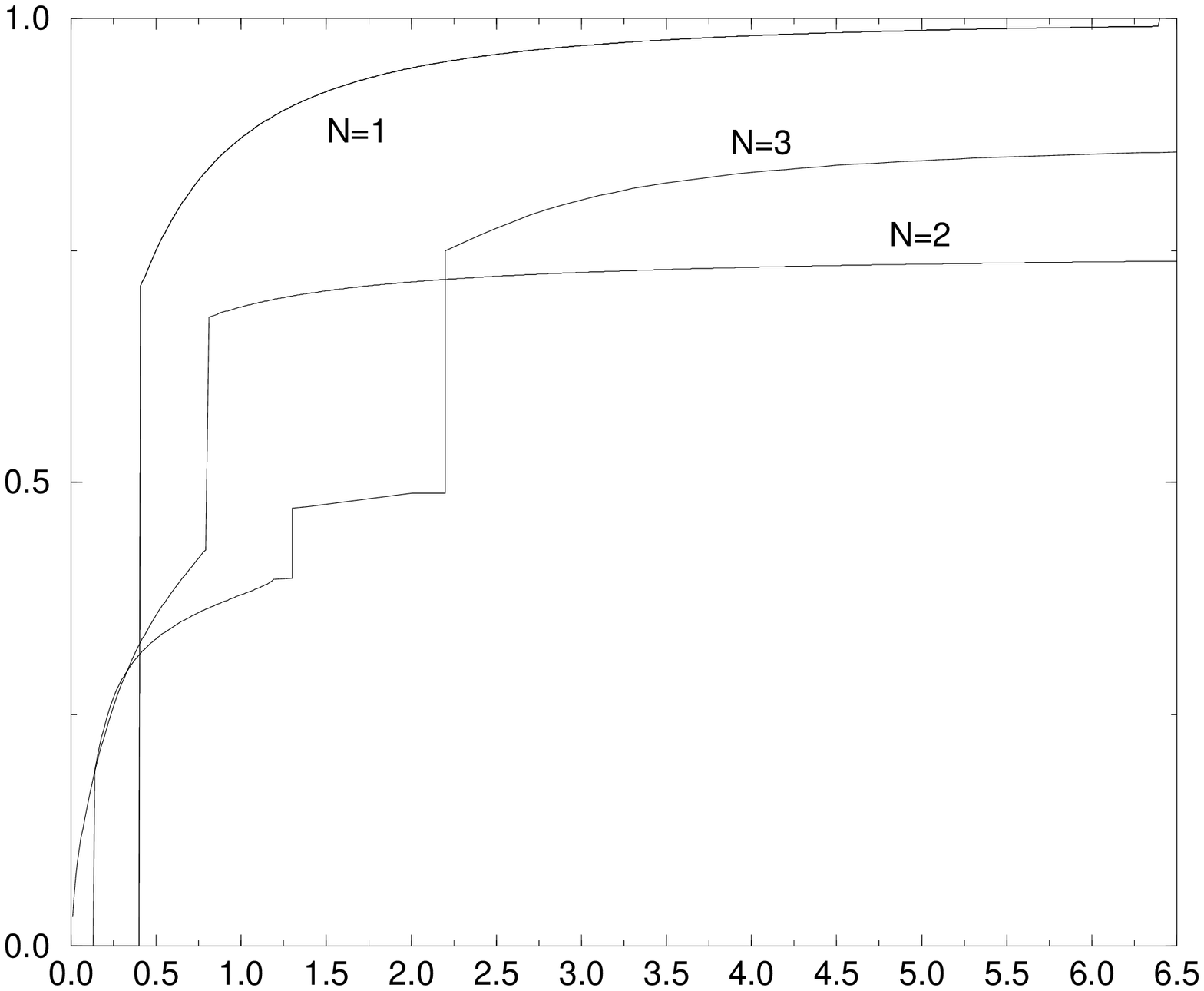,height=15cm,width=14cm,angle=270}}}
\end{picture}
\end{figure}


\begin{figure}[h]
\leavevmode
\centering
\setlength{\unitlength}{1cm}
\begin{picture}(15,21)
\put(6.5,21){\mbox{\Large\bf Fig.~4}} 
\put(-1,10.5){\makebox{\LARGE\boldmath $p_0(x)$}}
\put(6.5,-0.2){\makebox{\LARGE\boldmath $x$}}
\put(0,20){\makebox{\epsfig{file=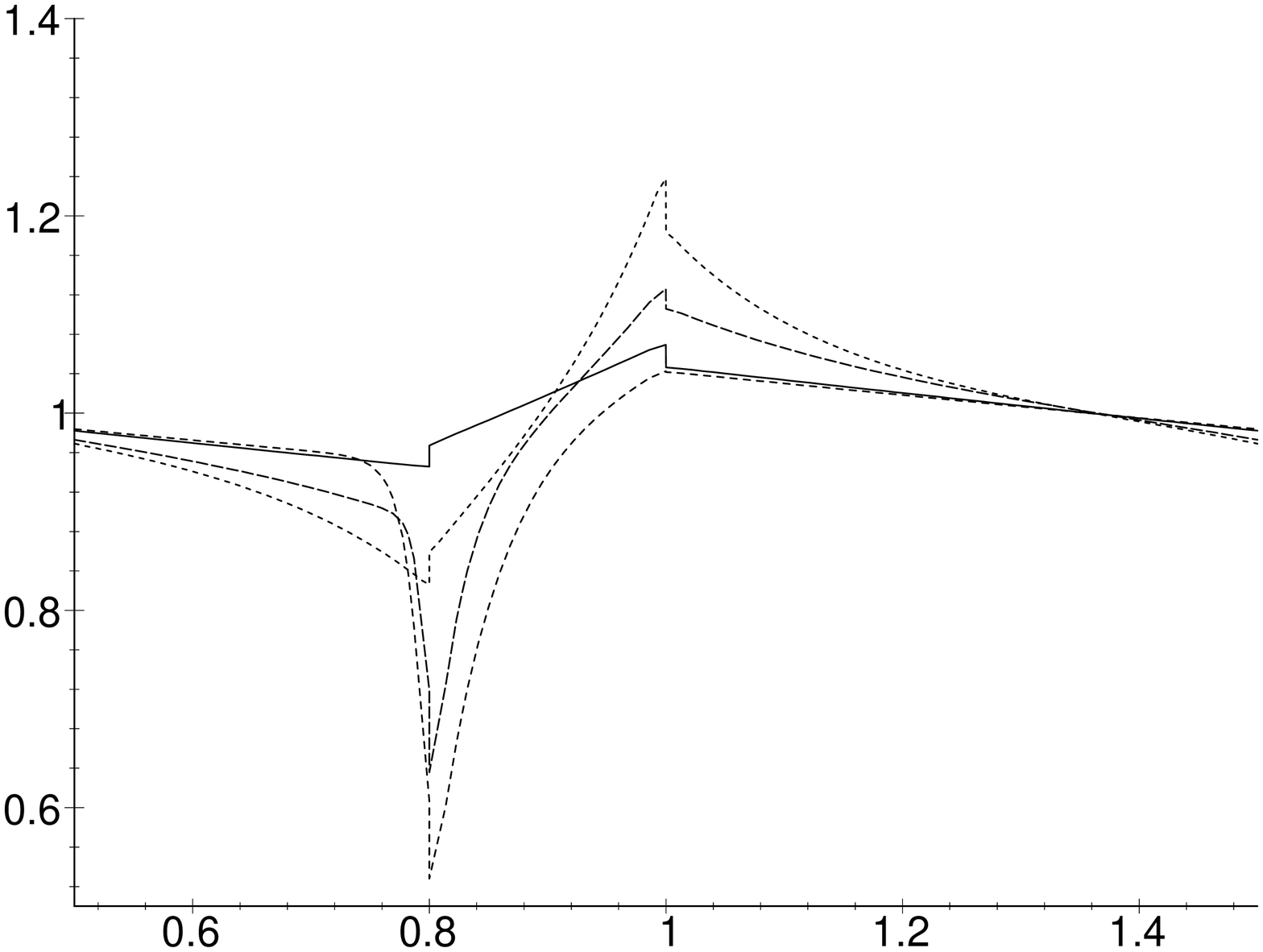,height=15cm,width=20cm,angle=270}}} 
\end{picture}
\end{figure}


\begin{figure}[h]
\leavevmode
\centering
\setlength{\unitlength}{1cm}
\begin{picture}(15,12)
\put(6.5,12){\mbox{\Large\bf Fig.~5a)}} 
\put(-0.5,10.5){\mbox{\large\boldmath $J/10^{-3}\Delta V$}}
\put(6,2){\mbox{\large\boldmath$\tau/10^{-3}\Delta V^{-1}$}}
\put(0,10.5){\mbox{
\epsfig{file=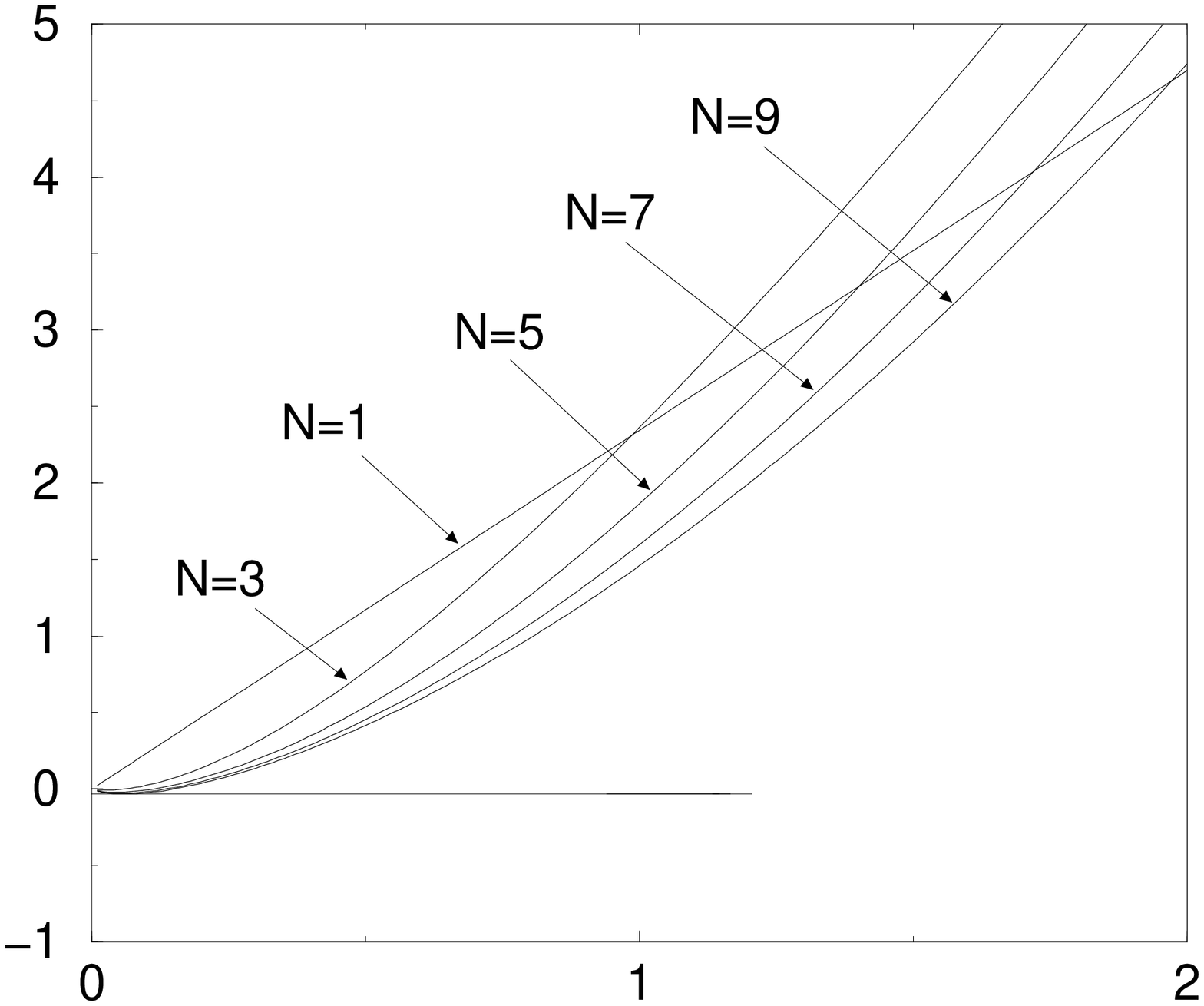,height=12.5cm,width=8cm,angle=270}}}
\put(6.8,6.7){\mbox{
\epsfig{file=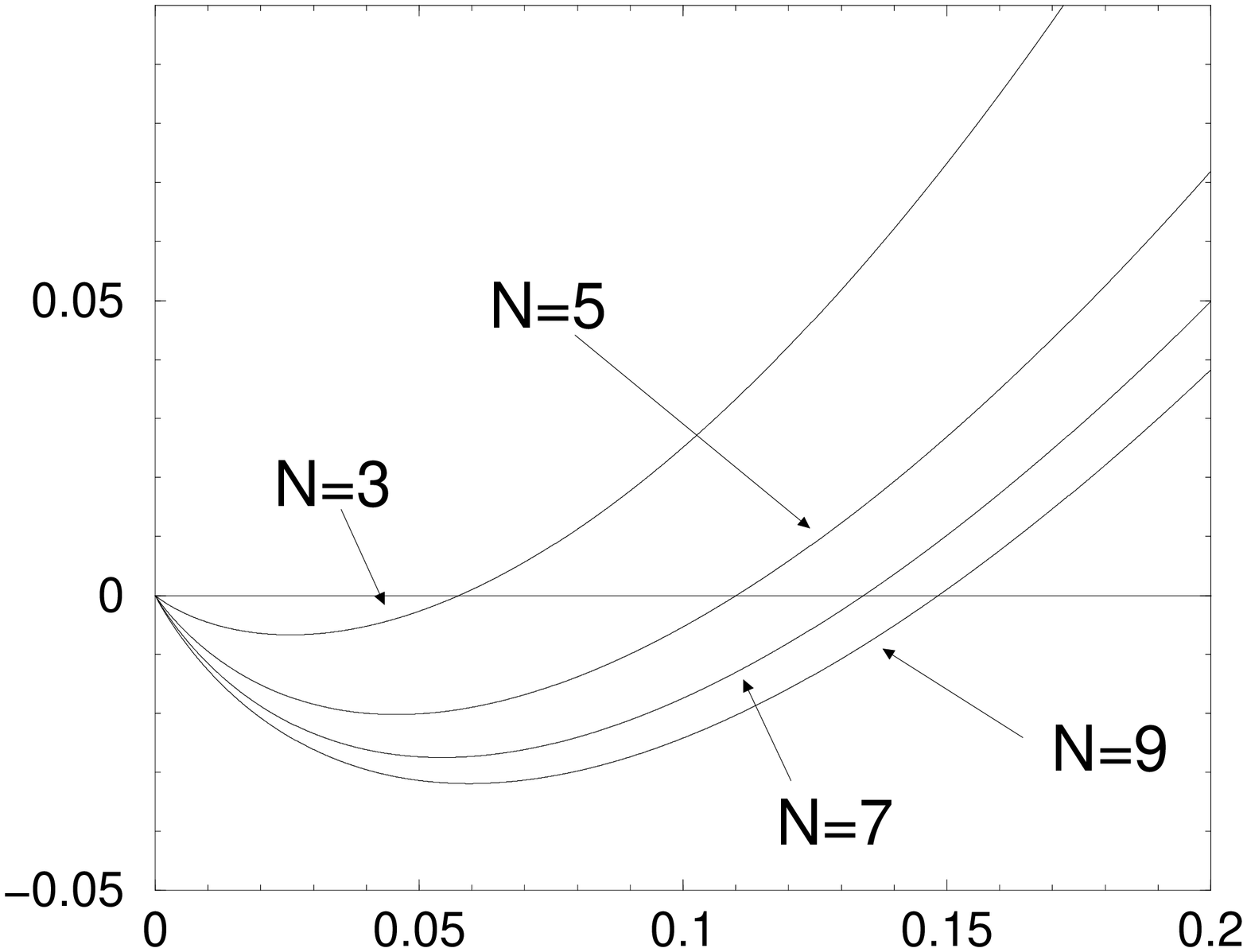,height=4.8cm,width=3.5cm,angle=270}}}
\end{picture}

\begin{picture}(15,12)
\put(6.5,12){\mbox{\Large\bf Fig.~5b)}} 
\put(-0.5,10.5){\mbox{\large\boldmath $J/\Delta V$}}
\put(6,2){\mbox{\large\boldmath$\tau/\Delta V^{-1}$}}
\put(0,10.5){\mbox{
\epsfig{file=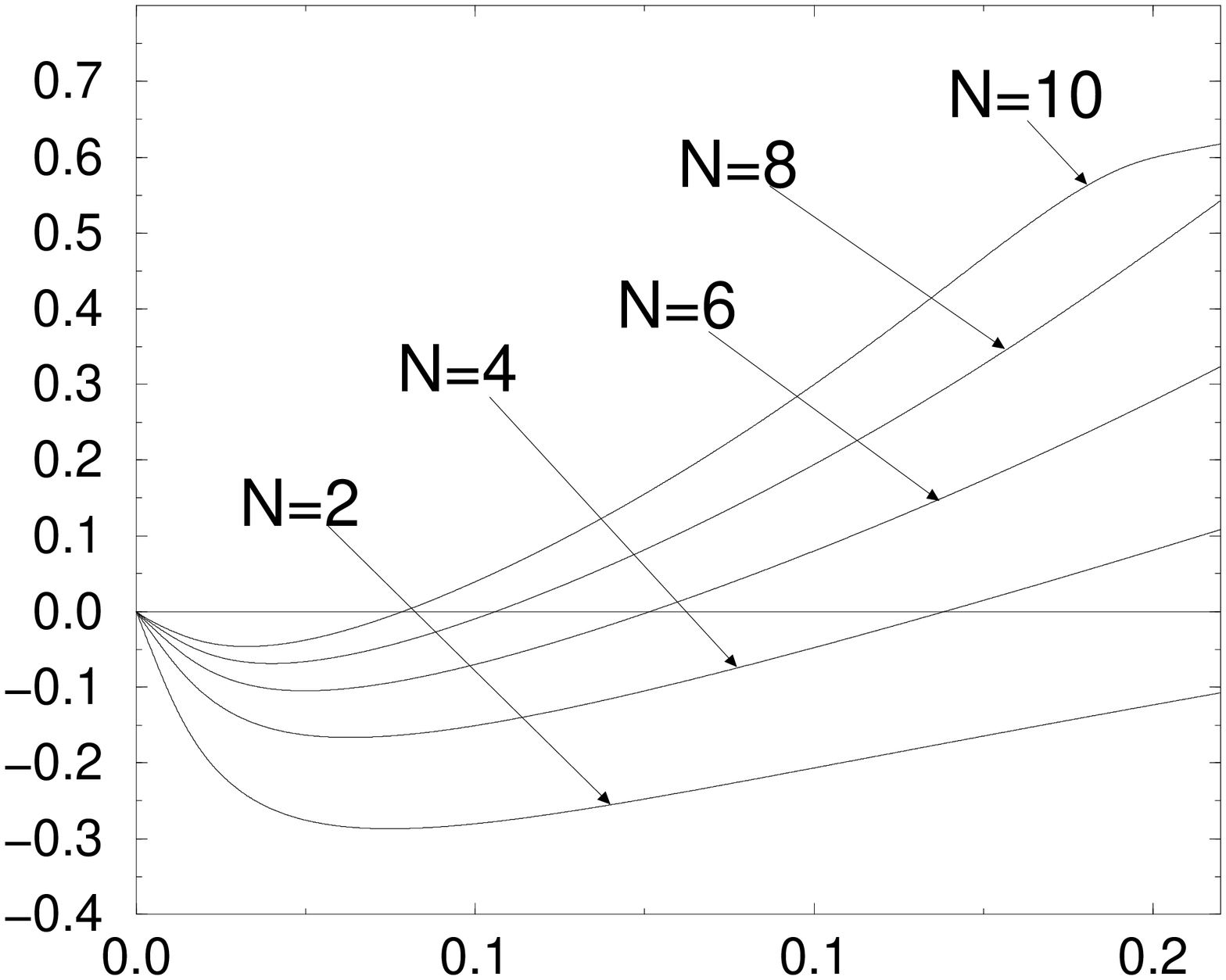,height=12.5cm,width=8cm,angle=270}}}
\end{picture}
\end{figure}


\begin{figure}[h]
\leavevmode
\centering
\setlength{\unitlength}{1cm}
\begin{picture}(15,21)
\put(6.5,21){\mbox{\Large\bf Fig.~6}} 
\put(0.5,10.5){\makebox{\large\boldmath $J$}}
\put(3.5,5.5){\makebox{\large\boldmath $K(2)$}}
\put(8.5,7.5){\makebox{\large\boldmath $N=2$}}
\put(9.5,13){\makebox{\large\boldmath $N=3$}}
\put(6.5,-0.2){\makebox{\large\boldmath $\tau/10^{-3}$}}
\put(0,20){\makebox{
\epsfig{file=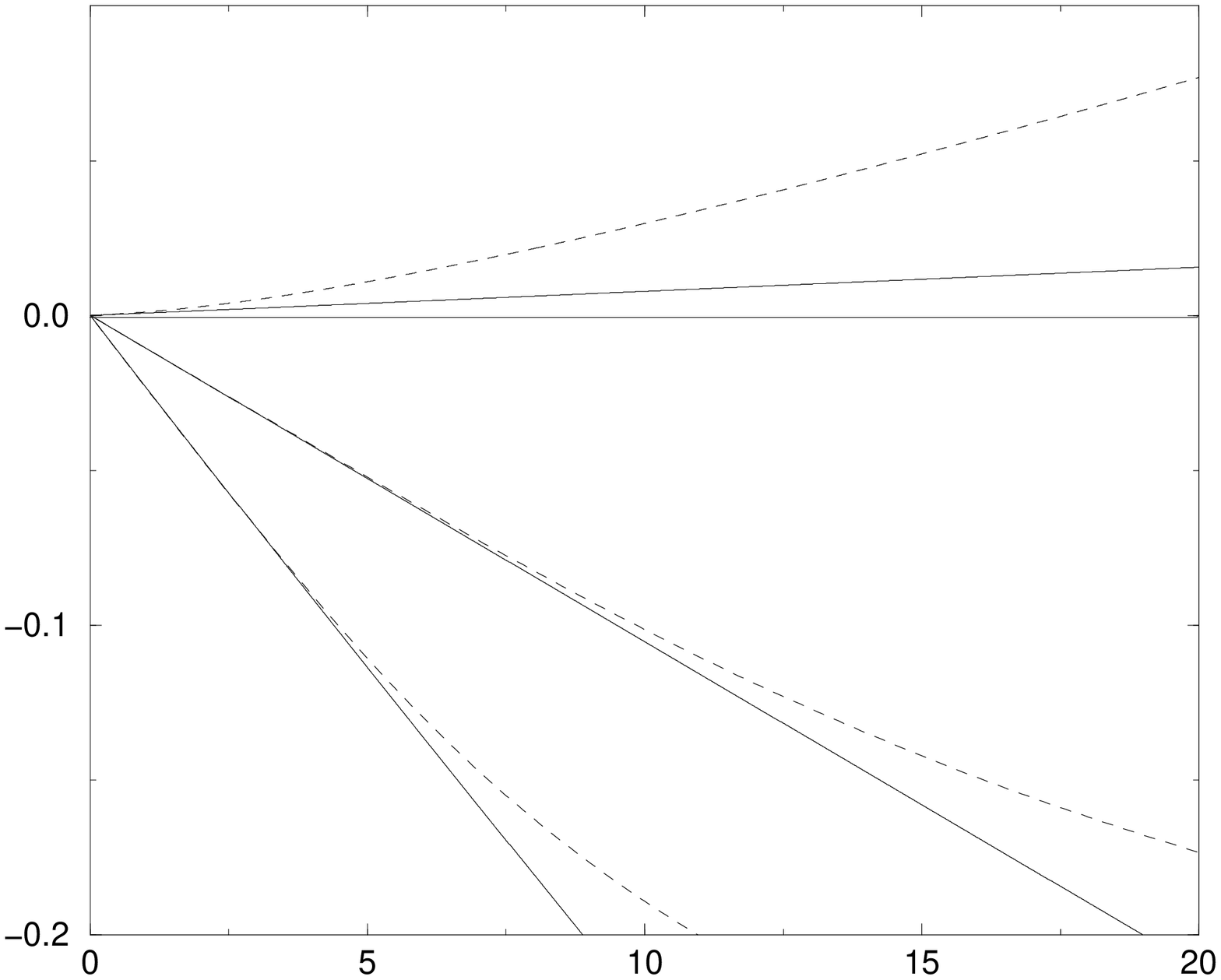,height=15cm,width=20cm,angle=270}}}
\end{picture}
\end{figure}

\end{document}